\documentclass[pra,aps,10pt,twocolumn,floatfix]{revtex4-1}

\usepackage{graphicx,color}
\graphicspath{{figs_paper/}}
\usepackage[nice]{nicefrac}							
\usepackage{amsmath,amssymb,bm}
\usepackage{braket}		

\usepackage[plainpages=false,pdfpagelabels,colorlinks=true,linkcolor=red,urlcolor=blue,citecolor=blue,pdftitle={},pdfauthor={},pdfdisplaydoctitle=true,pdfduplex=DuplexFlipLongEdge]{hyperref}

\definecolor{darkred}{rgb}{0.90,0.2,0.2}
\definecolor{darkgreen}{rgb}{0,0.60,.2}
\definecolor{darkblue}{rgb}{0.1,0.3,1}
\definecolor{grey}{cmyk}{0,0,0,0.25}
\definecolor{orange}{cmyk}{0,0.6,0.8,0}

\begin{document}

\title{Ergodicity Breaking Transition in Zero Dimensions}

\author{Jan \v Suntajs}
\author{Lev Vidmar}
\affiliation{Department of Theoretical Physics, J. Stefan Institute, SI-1000 Ljubljana, Slovenia}
\affiliation{Department of Physics, Faculty of Mathematics and Physics, University of Ljubljana, SI-1000 Ljubljana, Slovenia\looseness=-1}


\begin{abstract}
It is of great current interest to establish toy models of ergodicity breaking transitions in quantum many-body systems.
Here we study a model that is expected to exhibit an ergodic to nonergodic transition in the thermodynamic limit upon tuning the coupling between an ergodic quantum dot and distant particles with spin-1/2.
The model is effectively zero dimensional, however, a variant of the model was proposed by De Roeck and Huveneers to describe the avalanche mechanism of ergodicity breaking transition in one-dimensional disordered spin chains.
We show that exact numerical results based on the spectral form factor calculation accurately agree with theoretical predictions, and hence unambiguously confirm existence of the ergodicity breaking transition in this model.
We benchmark specific properties that represent hallmarks of the ergodicity breaking transition in finite systems.
\end{abstract}

\maketitle

{\it Introduction.} A fascinating property of isolated interacting quantum many-body systems is their ability to thermalize.
This statement usually refers to the properties of local observables~\cite{deutsch_91, srednicki_94, srednicki_99, rigol_dunjko_08, dalessio_kafri_16, Eisert2015, mori_ikeda_18, deutsch_18}, while many generic properties of ergodic systems can often be understood by analysing statistical properties of their energy spectra and comparing them to the predictions of random matrix theory~\cite{mehta_91, dalessio_kafri_16}.

Nevertheless, it was shown experimentally already more than 15 years ago that isolated interacting quantum many-body systems may not always thermalize~\cite{kinoshita_wenger_06}, at least on the experimentally relevant time scales~\cite{kinoshita_wenger_06, gring_kuhnert_12, langen15a, li_zhou_20, wilson_malvania_20}.
One class of interacting quantum systems exhibiting absence of thermalization and nonergodic dynamics are integrable systems, the spin-1/2 Heisenberg chain with translational invariance being a paradigmatic example~\cite{tomaz_quasilocal11, ilievski15, ilievski_medenjak_2016, bertini_heidrichmeisner_21}.
However, it is likely that in the thermodynamic limit, nonergodicity in these systems is  not robust against adding small integrability breaking terms~\cite{santos_04, barisic_prelovsek_09, torresherrera_santos_14, mierzejewski_prosen_15, bertini_essler_15, bertini_essler_16, mallayya_rigol_18, tang_kao_18, richter_jin_20, brenes_leblond_20}.
It is therefore of great scientific interest to uncover other classes of interacting quantum systems that exhibit robust counterexamples to thermalization.
Perhaps the most fascinating property of such systems would be the emergence of an ergodicity breaking phase transition between an ergodic and a nonergodic phase.

Recent experimental activities have led to demonstrations of different fingerprints of potentially robust nonergodic dynamics in interacting quantum systems at experimentally accessible times~\cite{schreiber_hodgman_15, smith_lee_16, choi_hild_16, lukin_rispoli_19, scherg_kohlert_21}.
From the theoretical perspective, however, it remains far from clear what are the universal properties of ergodicity breaking transitions, which are the relevant toy models, and what tools should one apply to detect the transition.

The latter statement may be illustrated by the example of a possible ergodicity breaking in spin-1/2 chains in random magnetic fields that act as quenched disorder.
It was proposed that such systems, which are ergodic at weak disorder, undergo an ergodicity breaking phase transition upon increasing the disorder~\cite{pal_huse_10}.
While the fate of this transition in the thermodynamic limit is still under debate~\cite{suntajs_bonca_20a, suntajs_bonca_20, sierant_lazo_21, leblond_sels_21, sels_polkovnikov_21, sels_polkovnikov_22, sels_22, kieferemmanouilidis_unanyan_20, luitz_barlev_20, kieferemmanouilidis_unanyan_21, ghosh_znidaric_22, schulz_taylor_20, sierant_delande_20, abanin_bardarson_21, corps_molina_21, hopjan_orso_21, herbrych_mierzejewski_22, crowley_chandran_22}, it was noticed that many properties of finite systems at rather large disorder appear to be nonergodic, however they may eventually become ergodic in the thermodynamic limit~\cite{suntajs_bonca_20a}.
This perspective has been then further supported using various different theoretical and numerical arguments~\cite{sels_polkovnikov_21, sels_polkovnikov_22, sels_22, vidmar_krajewski_21, morningstar_colmenarez_22, sierant_zakrzewski_22}.

\begin{figure}[b]
\centering
\includegraphics[width=0.90\columnwidth]{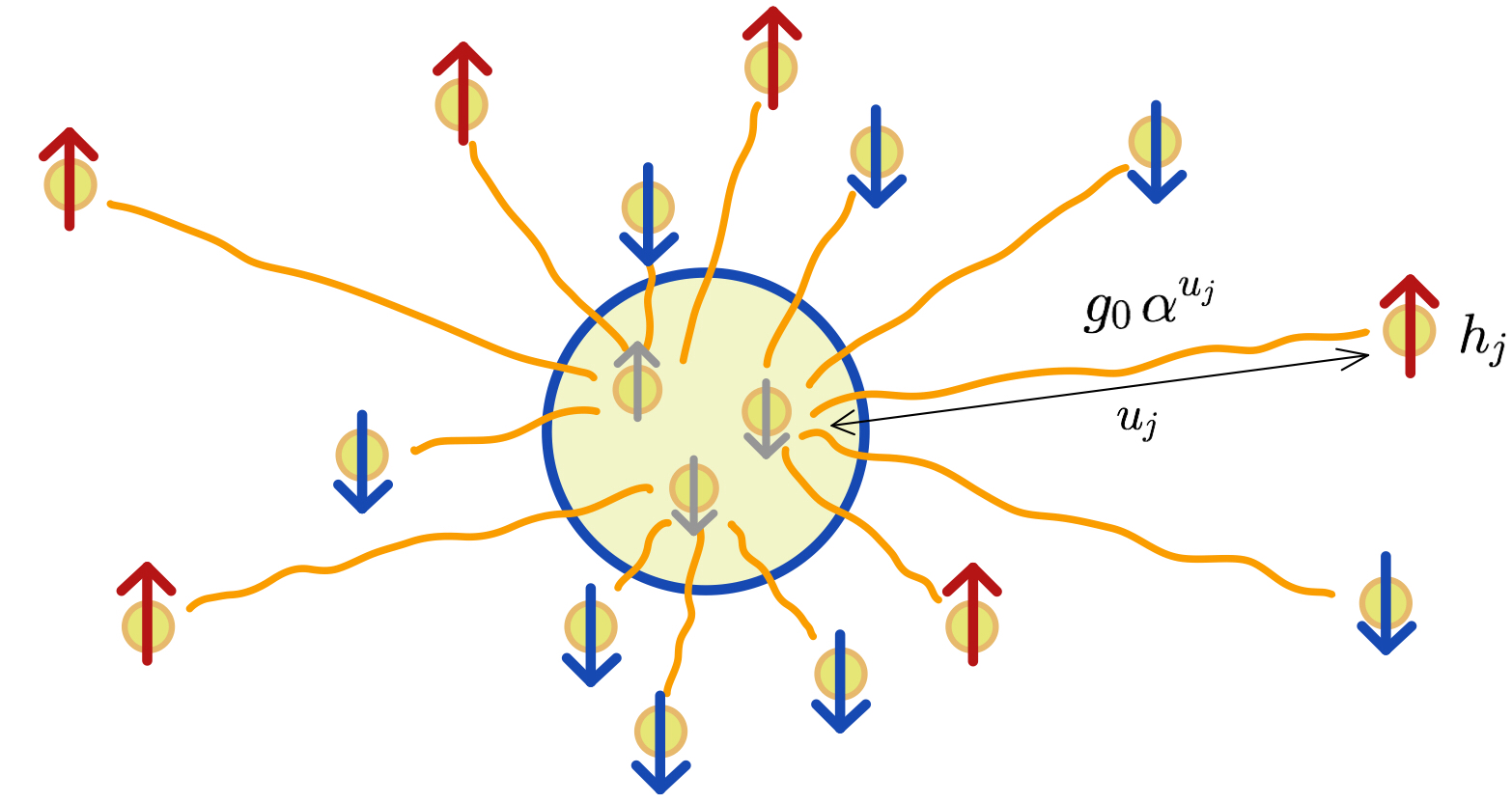}
\caption{
Sketch of the model from Eq.~(\ref{eq:def_model}).
Interactions of particles within the dot (blue circle) are described by a random matrix.
A particle $j$ outside the dot experiences the magnetic field $h_j$, and its distance from the dot is $u_j$.
The coupling amplitude between a particle $j$ and a randomly selected particle in the dot is $g_0 \alpha^{u_j}$.
}
\label{fig1}
\end{figure}

\begin{figure*}[!t]
\includegraphics[width=2.00\columnwidth]{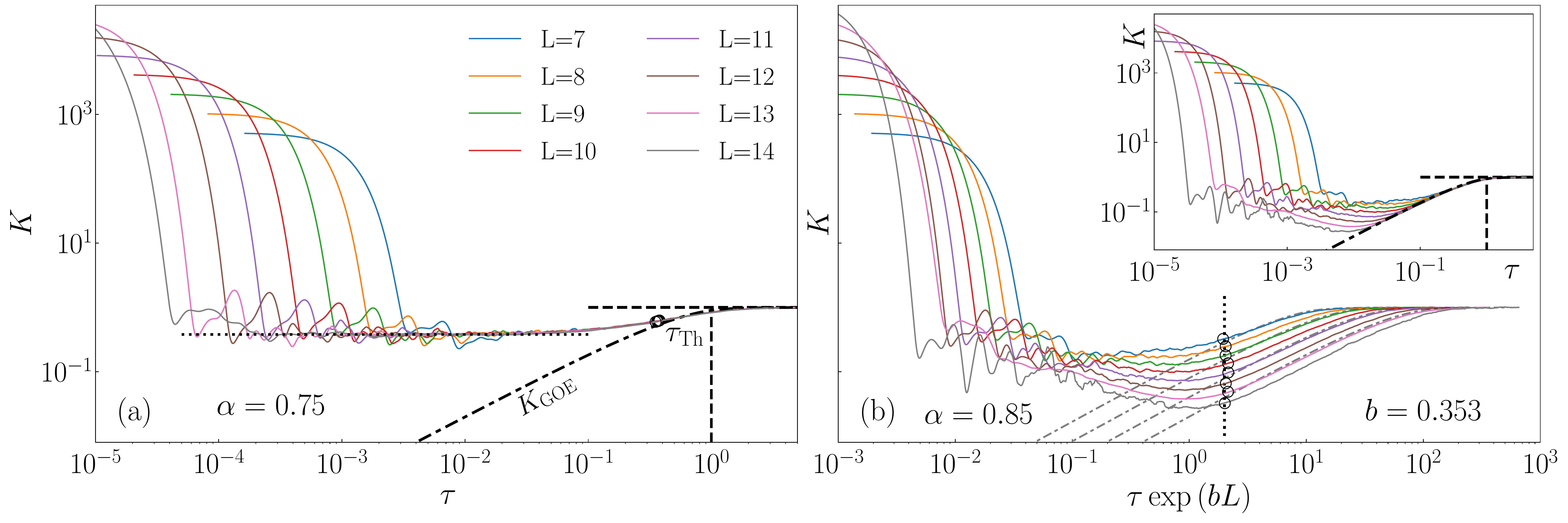}
\caption{
Spectral form factor $K(\tau)$ at different system sizes $L$.
Dashed-dotted lines denote the GOE results $K_{\rm GOE}(\tau) = 2\tau -\tau\ln(1+2\tau)$, open circles denote the extracted values of $\tau_{\rm {Th}}$, and the vertical dashed line is the scaled Heisenberg time $\tau_{\rm H}=1$.
(a) $K(\tau)$ at $\alpha=\alpha_c =0.75$.
The horizontal dotted line is the fit $K(\tau) \approx 0.38$.
(b) $K(\tau)$ at $\alpha=0.85$.
The inset shows the raw data while in the main panel we rescale $\tau \to \tau \exp(bL)$, where $b=0.353$, such that the rescaled values of $\tau_{\rm {Th}}$ coincide.
See~\cite{suppmat_sup} for a detailed description of the numerical extraction of $\tau_{\rm Th}$.
}
\label{fig2}
\end{figure*}

Based on the above arguments, it appears crucial to establish generic toy models of ergodicity breaking transitions, for which theoretical predictions and exact numerical results agree both qualitatively and quantitatively.
In this Letter, we achieve this goal for an interacting zero-dimensional model sketched in Fig.~\ref{fig1}, which is expected to exhibit features of the avalanche mechanism introduced by De Roeck and Huveneers~\cite{deroeck_huveneers_17}.
Based on the spectral form factor (SFF) analysis, we benchmark several properties of the ergodicity breaking transition, such as:
(a) signatures of a universal SFF shape at the transition point,
(b) exponential dependence of the Thouless time on both the system size and the interaction, which can be detected already deep in the ergodic regime,
and (c) extraction of a diverging length scale at the transition.
We argue that these properties may represent hallmarks of an ergodicity breaking transition (EBT) in finite systems.

{\it Model.}
We study a model of spin-1/2 particles that consists of two subsystems, as sketched in Fig.~\ref{fig1}:
a subsystem of $N$ particles with all-to-all interactions (referred to as a {\it dot}), and a subsystem of $L$ particles outside the dot, where each particle is only coupled to a single particle within the dot.
The full model Hamiltonian reads:
\begin{equation}\label{eq:def_model}
    \hat H = R + \sum\limits_{j=1}^{L} g_0\alpha^{u_j}\hat{S}^x_{n_j}\hat{S}_j^x + \sum\limits_{j=1}^{L} h_j \hat{S}_j^z.
\end{equation}
Properties of $N$ particles within the dot are described by a $2^N \times 2^N$ random  matrix $R$ drawn from the Gaussian orthogonal ensemble (GOE) that acts nontrivially on dot's degrees of freedom only.
The fields $h_j$ that act on particles outside the dot are drawn from a random box distribution, $h_j \in [0.5, 1.5]$.
In the coupling term, $\hat{S}^x_{n_j}$ acts on a randomly selected site $n_j$ within the dot, while $\hat{S}_j^x$ acts on the particle $j$ outside the dot.
The tuning parameter of the coupling strength $g_0\alpha^{u_j}$ is the parameter $\alpha$ (we set $g_0 \equiv 1$), while $u_j$ represents the distance between a coupled particle and the dot.
The latter is sampled from a random box distribution $u_j \in [j-\zeta_j, j+\zeta_j]$.
Apart from the energy conservation, the system has no other conservation laws and its corresponding total Hilbert space dimension equals $\mathcal{D}=2^{N + L}$.
We set $N=3$ and $\zeta_j=0.2$ throughout the study (see~\cite{suppmat_sup} for details).

A similar version of the model was used in Refs.~\cite{deroeck_huveneers_17, luitz_huveneers_17, crowley_chandran_20} in the description of the avalanche mechanism in one-dimensional (1D) strongly disordered spin chains, and a related model was studied in Ref.~\cite{ponte_laumann_17} to shed light on instability of nonergodicity in higher dimensions.
Here we argue that the model~\eqref{eq:def_model} is a zero-dimensional model since the thermodynamic limit is obtained by sending the number of particles outside the dot $L\to \infty$ while their coordination number $z=1$, and the number of particles within the dot $N$ are fixed.
Accordingly, we treat the model as a toy model to describe the EBT in zero dimensions and steer away from making any predictions regarding 1D (or higher-D) systems.

{\it Spectral form factor (SFF).}
The central quantity in our study is the SFF $K(\tau)$, which is the Fourier transform of the two-point spectral correlations, defined as
\begin{equation}\label{def_Kt}
\begin{split}
K(\tau) &= \frac{1}{Z} \left\langle \left|\sum_{n = 1}^{\cal D} \rho(\varepsilon_n) e^{-i 2\pi\varepsilon_n \tau}\right|^2 \right\rangle \, .
\end{split}
\end{equation}
Here,  $\{ \varepsilon_1\le\varepsilon_2\le \cdots\varepsilon_{\cal D} \}$ is the complete set of Hamiltonian eigenvalues after spectral unfolding, $\tau$ is the scaled time, and the average $\langle ... \rangle$ is carried out over different realizations of $\hat H$.
We follow the implementation of $K(\tau)$ from~\cite{suntajs_bonca_20a}, which is for convenience summarized in the Supplemental Material~\cite{suppmat_sup} (the normalization $Z$ and a smooth filtering function $\rho(\varepsilon_n)$ that eliminates contributions of the spectral edges are provided there).
Numerically, we study systems with up to $L + N = 17$ sites, thus requiring full exact diagonalization of matrices up to dimension $\mathcal{D} \times \mathcal{D},$ where $\mathcal{D} = 2^{17} = 131072$.

Our main intuition for why the SFF should represent a useful tool to detect the EBT is based on the analysis of the localization transition in the three-dimensional (3D) Anderson model~\cite{suntajs_prosen_21}, for which the transition point is known to high accuracy~\cite{slevin_ohtsuki_18, suntajs_prosen_21}.
At the transition point in the 3D Anderson model, $K(\tau)$ exhibits a universal shape that is independent of the system size for a wide interval of $\tau$, and it consists of two regimes~\cite{suntajs_prosen_21}:
$K(t) \approx {\rm const}$ at $\tau \ll 1$, followed by a short interval around $\tau \lesssim 1$ where $K(\tau) \approx K_{\rm GOE}(\tau) = 2\tau -\tau\ln(1+2\tau)$.
We define the Thouless time $\tau_{\rm Th}$ (in scaled units) as the time when the SFF $K(\tau)$ approaches the GOE prediction $K_{\rm GOE}(\tau)$.
If $K(\tau)$ is universal and independent of system size $L$, the same holds true for $\tau_{\rm Th}$.
One can hence consider independence of $\tau_{\rm Th}$ on $L$ as a criterion for the transition.

Remarkably, a very similar structure of $K(\tau)$ is also observed in the model from Eq.~(\ref{eq:def_model}), which is shown in Fig.~\ref{fig2}(a) at $\alpha=0.75$.
At this value of $\alpha$, $K(\tau)$ appears to exhibit the most $L$-independent form [apart from the short time limit $\tau \to 0$ when $K(\tau)\gg 1$].
The value of $K(\tau)$ in the broad $\tau$-independent regime is $K(\tau) \approx 0.38$, which is very close to the one in the 3D Anderson model at the transition point (cf.~Fig.~8(a) in~\cite{suntajs_prosen_21}).

Before proceeding, we note that our results suggest the EBT to occur at $\alpha \approx 0.75$, and hence we set $\alpha_c = 0.75$ further on.
Some analytical arguments (to be presented below and also in~\cite{deroeck_huveneers_17, luitz_huveneers_17}) predict the transition to occur at $\bar\alpha=1/\sqrt{2}\approx 0.71$. While our numerical results are not inconsistent with the latter prediction, we also refrain from making any sharp predictions of the transition point in systems much larger than those studied here, or in systems with a different choice of model parameters.

Figure~\ref{fig2}(b) displays results for $K(\tau)$ in the ergodic regime at $\alpha=0.85$.
In the inset we show that $\tau_{\rm Th} \to 0$ in the thermodynamic limit $L\to\infty$, which can be seen as a hallmark of ergodicity.
The most remarkable property of $K(\tau)$ can be observed when the scaled time $\tau$ is multiplied by a factor $\exp(bL)$, where $b$ is a constant.
Results shown in the main panel of Fig.~\ref{fig2}(b) suggest that after this rescaling, the Thouless time becomes nearly independent of $L$.
This indicates that the Thouless time in physical units scales exponentially with $L$ in the ergodic regime, and will be analysed in more detail below.

\begin{figure}[!t]
\begin{center}
\includegraphics[width=0.99\columnwidth]{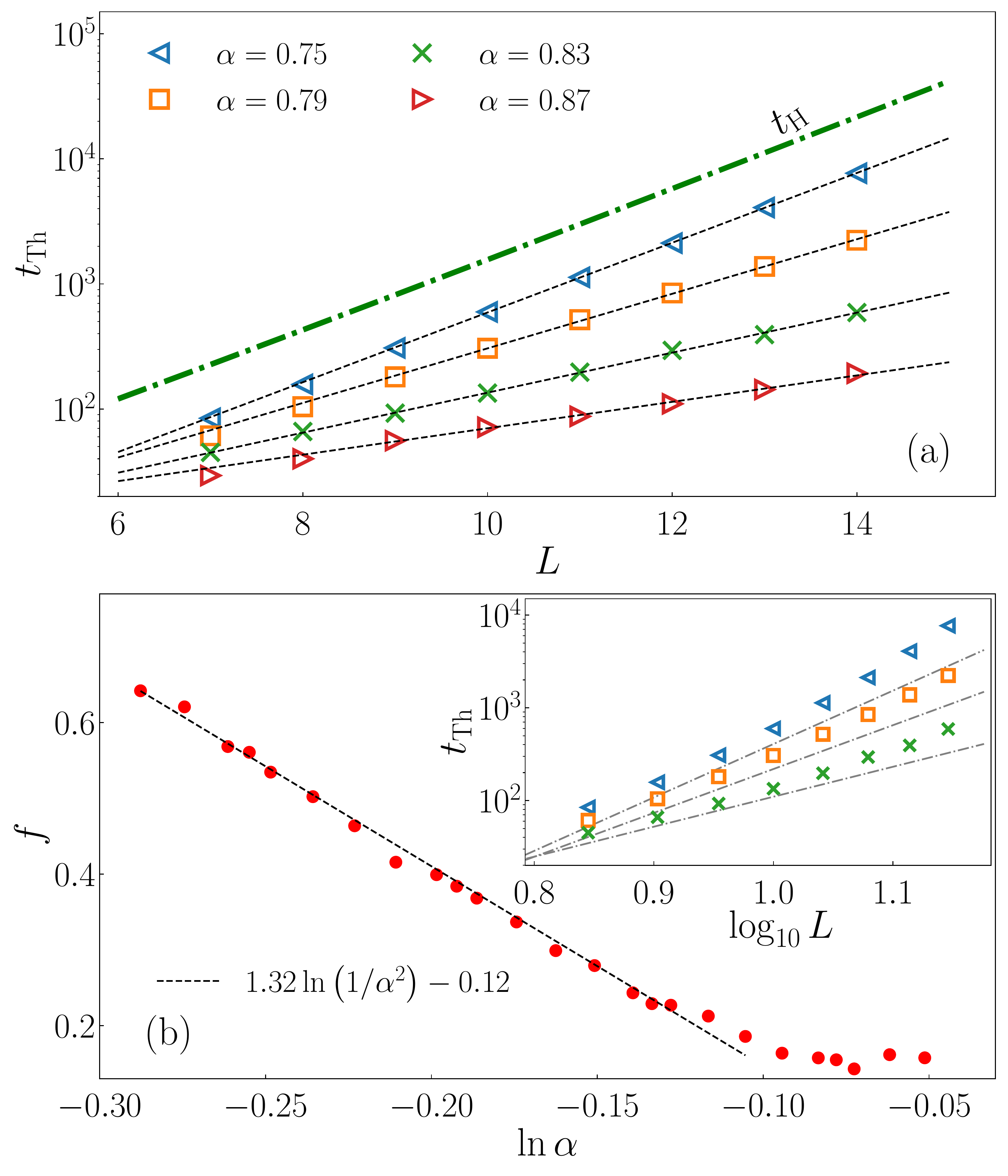}
\end{center}
\caption{
(a) Scaling of the Thouless time $t_{\rm Th}$ with $L$ on a log-linear scale for different values of $\alpha$ (symbols).
Dashed lines are fits to Eq.~(\ref{eq:def_tTh}), while the dashed-dotted line is the scaling of the Heisenberg time $t_{\rm H}$.
The same results are shown in the inset of (b) on a log-log scale, where dashed-dotted lines are guides to the eye with the same slope as the power-law fits to the numerical data.
(b) Dependence of $f$ on $\ln \alpha$, where $f$ is obtained by fitting Eq.~\eqref{eq:def_tTh} to the results for $t_{\rm Th}(L)$.
Line is the fitted function $f(\alpha)=1.32\ln\left(1/\alpha^2\right) - 0.12$. 
} \label{fig3}
\end{figure} 

{\it Thouless time in the ergodic regime.}
From now on we consider the Thouless time in physical units, which is defined as $t_{\rm{Th}} = \tau_{\rm{Th}} \hbar / \overline{\delta E}$, where $\overline{\delta E}$ is the mean level spacing~\cite{suppmat_sup} and we set $\hbar \equiv 1$.
In Fig.~\ref{fig3}(a), we show $t_{\rm Th}$ versus $L$ in a log-linear scale at different $\alpha$ in the ergodic regime, and at $\alpha=\alpha_c$.
The results suggest that $t_{\rm Th}$ increases exponentially with $L$, at least in the regime $\alpha \lesssim 0.87$.
This is corroborated in the inset of Fig.~\ref{fig3}(b) where the same results are shown on a log-log scale, exhibiting a growth with $L$ that is faster than power-law.
We describe these results with the ansatz
\begin{equation}\label{eq:def_tTh}
t_{\rm Th} \propto e^{f L} \;.
\end{equation}
An important property of this ansatz is that $f=f(\alpha)$ is a function of the coupling strength, as evident from the varying slopes of $t_{\rm Th}(L)$ in a log-linear scale in Fig.~\ref{fig3}(a).
The dashed-dotted line in Fig.~\ref{fig3}(a) denotes the scaling with $L$ of the Heisenberg time $t_{\rm H} = 1/\overline{\delta E}$, which has the same slope as $t_{\rm Th}(L)$ at $\alpha=\alpha_c$.

We can explain the relation in Eq.~(\ref{eq:def_tTh}) by a rather crude and simple approximation, which is however accurate enough for building our intuition about the key scaling relations in the system.
Since $t_{\rm Th}$ may be seen as the longest physically relevant time scale, we estimate its inverse, denoted $E_{\rm Th}^*$, by the coupling of the farthermost particle to the dot, $g_L = \alpha^L$.
Using the Fermi golden rule arguments, one gets 
$E_{\rm Th}^* = g_L^2/\varepsilon$, where $\varepsilon=\mathcal{O}(1)$~\cite{luitz_huveneers_17}.
Then, $t_{\rm Th}^* = 1/E_{\rm Th}^*$ is given by
\begin{equation}\label{eq:def_tThstar}
    t_{\rm Th}^* \propto \alpha^{-2L} = e^{\ln\left(\frac{1}{\alpha^2}\right)\,L}.
\end{equation}
In the main panel of Fig.~\ref{fig3}(b) we plot the rates $f$, obtained by fitting Eq.~(\ref{eq:def_tTh}) to the actual numerical results, versus $\ln \alpha$.
We fit the ansatz $f(\alpha) = a_1\ln\left(\frac{1}{\alpha^2}\right) + a_0$ to the results in the interval $\alpha \in [0.75, 0.87]$, and get $a_1 = 1.32$, $a_0 = -0.12$.
This is indeed reasonably close to the prediction by Eq.~(\ref{eq:def_tThstar}), $f(\alpha) = \ln\left(\frac{1}{\alpha^2}\right)$, which assumes $a_1=1$ and $a_0=0$.

\begin{figure}[!t]
\begin{center}
\includegraphics[width=0.99\columnwidth]{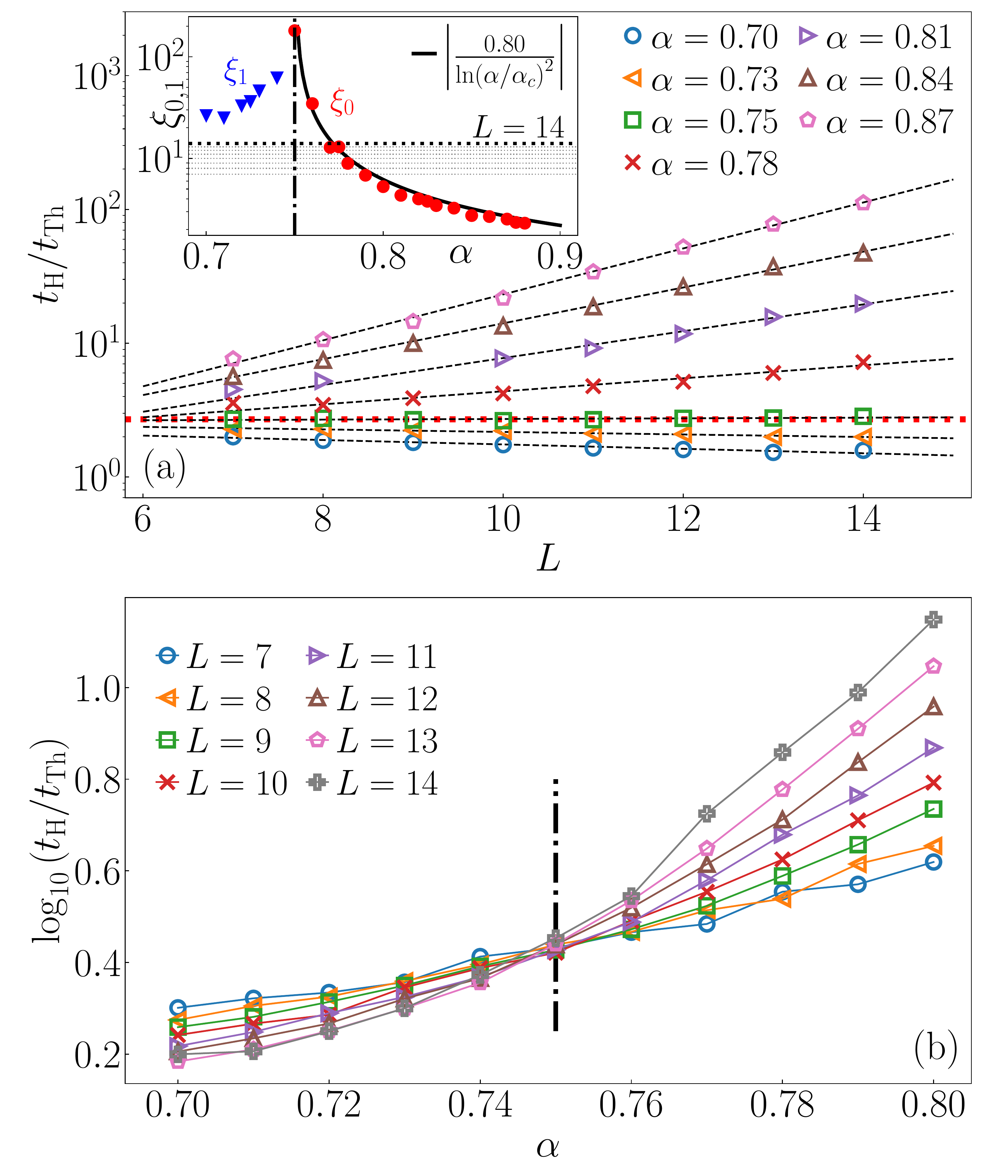}
\end{center}
\caption{
(a)
The ratio $t_{\rm H}/t_{\rm Th}$ versus $L$ at different $\alpha$.
Symbols are numerical results, while lines are fits of the exponential function from Eq.~(\ref{eq:def_GL2}), where $A$ and $\xi_0$ ($\xi_1$) are fitting parameters.
Inset: characteristic lengths $\xi_0$, $\xi_1$, extracted from the fits in the main panel (symbols).
Solid line is a one-parameter fit of a function $c_1/\ln(\alpha/\alpha_c)^2$ to $\xi_0$ in the ergodic regime, where we set $\alpha_c=0.75$ (vertical dashed-dotted line) and obtain $c_1=0.8$ from the fit.
The horizontal dashed line denotes $L=14$, i.e., the largest attainable system size.
(b) $\log_{10}(t_{\rm H}/t_{\rm Th})$ versus $\alpha$ at different $L$.
Vertical line denotes $\alpha_c=0.75$.
\label{fig4}}
\end{figure}

{\it Criterion for the EBT.} We next discuss the connection of the results presented so far with the emergence of EBT.
We expect that the system is ergodic when the Thouless time $t_{\rm Th}$ increases slower than the Heisenberg time $t_{\rm H} \propto 2^L = e^{\ln(1/\bar\alpha^2) L}$, where $\bar\alpha = 1/\sqrt{2}$, while the onset of nonergodic behavior occurs when both times scale identically.
This gives rise to the criterion for the EBT
\begin{equation} \label{def_tH_tTh}
    \frac{t_{\rm H}}{t_{\rm Th}} = {\rm const}.
\end{equation}
As argued in our discussion of Fig.~\ref{fig2}(a), this criterion in the 3D Anderson model is a natural consequence of the universal, $L$-independent shape of its SFF at the transition.
As argued in~\cite{suppmat_sup}, the criterion from Eq.~(\ref{def_tH_tTh}) is also consistent with the hybridization condition as a criterion for the EBT as used, e.g., in~\cite{deroeck_huveneers_17}.

The criterion from Eq.~(\ref{def_tH_tTh}) is tested in Figs.~\ref{fig4}(a) and~\ref{fig4}(b) for the numerically extracted values of $t_{\rm Th}$ and $t_{\rm H}$.
They suggest the EBT to occur at $\alpha\approx \alpha_c = 0.75$, at which $t_{\rm H}/t_{\rm Th}$ as a function of $L$ is nearly constant [see Fig.~\ref{fig4}(a)], and the curves for $t_{\rm H}/t_{\rm Th}$ at different $L$, plotted as a function of $\alpha$, cross at $\alpha=\alpha_c$ [see Fig.~\ref{fig4}(b)].
These results are consistent with the universal, $L$-independent shape of the SFF $K(\tau)$ at $\alpha=\alpha_c$ being the hallmark of the EBT, see Fig.~\ref{fig2}(a).

Considering Eq.~(\ref{eq:def_tThstar}) as a relevant approximation of $t_{\rm Th}$, one can express the criterion from Eq.~(\ref{def_tH_tTh}) as
\begin{equation}\label{eq:def_GL2}
    \frac{t_{\rm H}}{t_{\rm Th}} = \left\{
    \begin{array}{cc}
      A e^{\frac{L}{\xi_{0}}}\;,\;\; & \alpha>\bar\alpha \\
      A e^{-\frac{L}{\xi_{1}}}\;,\;\;   & \alpha<\bar\alpha
    \end{array}
    \right.\;,
\end{equation}
where the characteristic lengths $\xi_0$ and $\xi_1$ in the ergodic ($\alpha>\bar\alpha)$ and nonergodic ($\alpha<\bar\alpha)$ regimes, respectively, are defined as
\begin{equation}\label{eq:def_xi}
    \xi_0 = \frac{1}{\ln\left(\frac{\alpha}{\bar\alpha}\right)^2} \;,\;\;\;
    \xi_1 = \frac{1}{\ln\left(\frac{\bar\alpha}{\alpha}\right)^2} \;.
\end{equation}
One may think of an inverse of the characteristic lengths, say $1/\xi_0$, as the difference $1/\xi_0 = 1/\xi_{\rho} - 1/\xi_{\rm c}$, where $\xi_\rho = 1/\ln (1/\bar\alpha^2)$ sets the decay of the mean level spacing and $\xi_{\rm c} = 1/\ln (1/\alpha^2)$ sets the decay of the weakest coupling to the dot.

We fit the exponential function from Eq.~(\ref{eq:def_GL2}), where $A$ and $\xi_0$ ($\xi_1$) are fitting parameters, to the numerical results in the main panel in Fig.~\ref{fig4}(a).
The extracted characteristic lengths $\xi_0$ and $\xi_1$ indeed exhibit a tendency to diverge at the transition point $\alpha_c$.
We fit the characteristic length $\xi_0$ on the ergodic side with a function $c_1/\ln(\alpha/\alpha_c)^2$, where $\alpha_c=0.75$, and obtain $c_1=0.80$.
These results are fairly close to the prediction from Eq.~(\ref{eq:def_xi}), which assumes $\alpha_c=\bar\alpha=0.71$ and $c_1=1$.
We note that despite a reasonably good agreement between numerical results and analytical considerations, the latter may be refined in several ways, which we discuss in~\cite{suppmat_sup} in more detail.

{\it Conclusions.} In this Letter, we analysed the model that we suggest to be the toy model of the EBT in a zero-dimensional interacting system.
It exhibits three hallmarks of the EBT.
The first is a universal, system-size independent form of the SFF at the transition point, which strongly resembles the single-particle SFF of the 3D Anderson model at the localization transition.
Determining the analytical form of the universal SFF at the transition, as well as understanding the origin of agreement between the noninteracting 3D Anderson model and the interacting zero-dimensional model studied here, remains an open question for future research.

Another important feature is the exponential scaling of $t_{\rm Th}$ in the ergodic regime with the system size $L$, where the rate is a function of the interaction that drives the EBT.
In fact, this scaling is observed already deep in the ergodic regime where the level spacing ratio of adjacent gaps $r$ (see~\cite{suppmat_sup}) agrees with the GOE prediction, $r \approx 0.53$.
To our best knowledge, such a scaling of $t_{\rm Th}$ has so far not been observed in interacting disordered systems in dimension one or higher.

Finally, both analytical arguments and numerical results for the criterion of the EBT show a divergent characteristic length at the transition.
This length is, in the vicinity of the transition, much larger than the numerically accessible system sizes $L$.
In the future work it may be instructive to study other measures of the transition that are not based on spectral properties, and to explore their common features.

\acknowledgements
We acknowledge discussions with W. De Roeck and T. Prosen.
This work is supported by the Slovenian Research Agency (ARRS), Research core fundings No.~P1-0044 (J.\v S. and L.V.) and No.~J1-1696 (L.V.). We gratefully acknowledge the High Performance Computing Research Infrastructure Eastern Region (HCP RIVR) consortium \href{https://www.hpc-rivr.si/}{(www.hpc-rivr.si)} and The European High Performance Computing Joint Undertaking (EuroHPC JU) \href{https://eurohpc-ju.europa.eu/}{(eurohpc-ju.europa.eu)} for funding this research by providing computing resources of the HPC system Vega at the Institute of Information sciences \href{https://www.izum.si/en/home/}{(www.izum.si)}.

\bibliographystyle{biblev1}
\bibliography{references1,references2}

\newpage
\phantom{a}
\newpage
\setcounter{figure}{0}
\setcounter{equation}{0}
\setcounter{table}{0}

\renewcommand{\thetable}{S\arabic{table}}
\renewcommand{\thefigure}{S\arabic{figure}}
\renewcommand{\theequation}{S\arabic{equation}}
\renewcommand{\thepage}{S\arabic{page}}

\renewcommand{\thesection}{S\arabic{section}}

\onecolumngrid

\begin{center}

{\large \bf Supplemental Material:\\
Ergodicity Breaking Transition in Zero Dimensions}\\

\vspace{0.3cm}

\setcounter{page}{1}

Jan \v Suntajs and Lev Vidmar\\
{\it Department of Theoretical Physics, J. Stefan Institute, SI-1000 Ljubljana, Slovenia and} \\
{\it Department of Physics, Faculty of Mathematics and Physics, University of Ljubljana, SI-1000 Ljubljana, Slovenia}

\end{center}

\vspace{0.6cm}

\twocolumngrid

\label{pagesupp}

\section{Details about the numerical implementation}

In Eq.~\eqref{eq:def_model}, we model the interactions of $N$ particles within the dot by a random matrix $R$. We draw the latter from a GOE ensemble, $R=\frac{\beta}{2}\left(A + A^T\right) \in 2^N \times 2^N,$ where the matrix elements $A_{i, j}$ are sampled from a normal distribution with zero mean and unit variance, and $\beta=0.3$. As stated in the main text, we keep $N=\mathcal{O}(1)$ constant, setting $N=3$ in our numerical calculations.
The results shown in Fig.~\ref{fig2} are calculated using $N_\mathrm{samples}=1000$ different Hamiltonian realizations.
Otherwise, we use $N_\mathrm{samples}=1000$ for $L\leq 13$ and $N_\mathrm{samples}\approx 100$ for $L=14$ for all the results shown in the main text.

{\it Calculation of the Heisenberg time.} The Heisenberg time $t_{\rm H}$ is defined as the inverse of the mean level spacing, $t_{\rm H} = \hbar/\overline{\delta E},$ where $\hbar \equiv 1.$ We define the mean level spacing $\overline{\delta E} = \Gamma_0 / \left(\chi \mathcal{D}\right),$ where
\begin{equation}\label{eq:def_gamma0}
    \Gamma_0^2 = \left\langle \rm{Tr}\{\hat{H}^2\} / \mathcal{D} - \rm{Tr}\{\hat H\}^2/\mathcal{D}^2 \right\rangle
\end{equation}
is the variance after the averaging $\langle ... \rangle$ over different realizations of $\hat H$, $\mathcal{D}=2^{L + N}$, and $\chi$ controls the number of energy levels in the interval $[\bar E, \bar E + \Gamma_0],$ where $\bar E = \langle \rm{Tr} \{\hat H\}\rangle / \mathcal{D}.$
Assuming Gaussian density of states, we calculate $\chi$ as $\chi = \int_0^{\Gamma_0}\exp\{-E^2/(2\Gamma_0^2)\}/(\sqrt{2\pi}\Gamma_0)\rm{d}E,$ yielding $\chi = \left(1/2\right)\rm{erf}[1/\sqrt{2}] \approx 0.3413.$
We verified that numerically calculated values of $\Gamma_0^2$ match very well with the predictions for $\Gamma_0^2$ in the grandcanonical ensemble,
$\Gamma_0^2 = 2^{N - 1} \beta^2 + \left(\frac{g_0}{4}\right)^2 \frac{\alpha^{2L + 2} - \alpha^2}{\alpha^2 - 1} + \frac{L}{4}\left(1 + \frac{W^2}{3}\right)$. Noting that, for $\alpha < 1,$ only the last term in the expression for $\Gamma_0^2$ is extensive and the first two contributions are small, and hence we get
\begin{equation} \label{def_tH_auppmat}
    t_{\rm H} = \frac{\chi\mathcal{D}}{\Gamma_0} \propto \exp\{L\ln 2 - \frac{1}{2}\ln L\}.
\end{equation}
In the numerical calculations we use exact values of $t_{\rm H}$ as given by the left part of Eq.~(\ref{def_tH_auppmat}), while in the analytical considerations we only use the dominant $L-$dependent contribution $t_{\rm H} \propto \exp\{L \ln 2\}$.

\begin{figure*}[!]
\centering
\includegraphics[width=2\columnwidth]{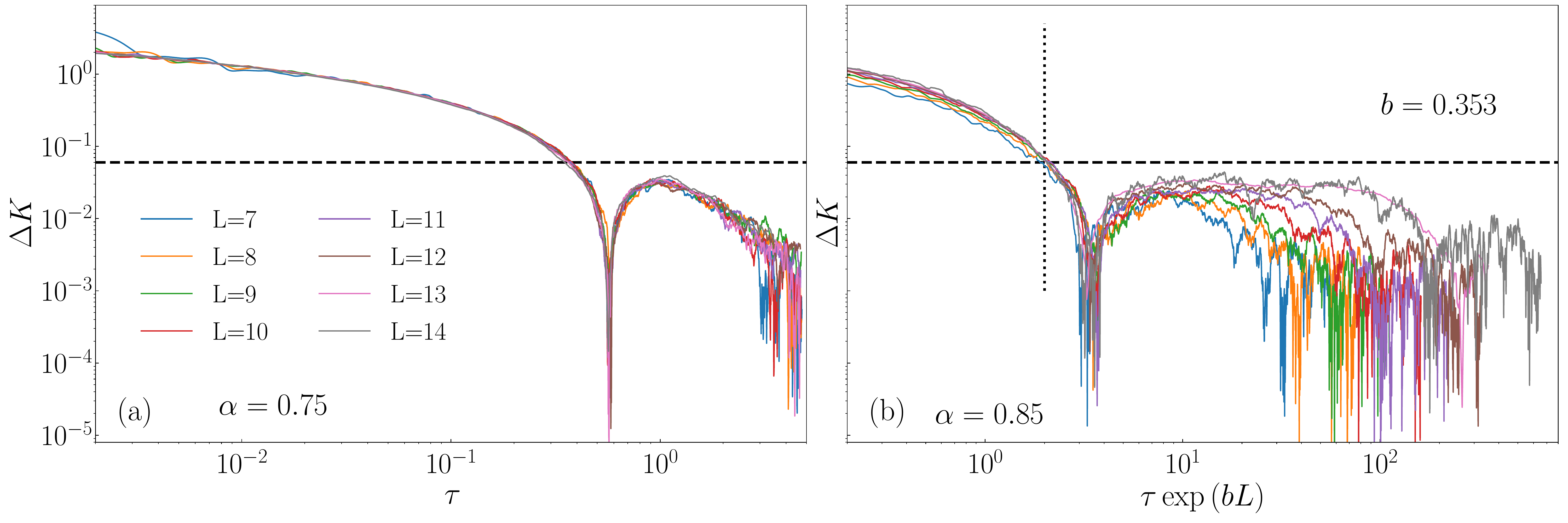}
\caption{Deviation measure $\Delta K(\tau)$ given by Eq.~\eqref{eq:def_deviation_measure} showing the approach of the numerically calculated $K(\tau)$ curves towards the GOE prediction $K_{\rm GOE}(\tau)$ for the same datasets as the ones shown in Fig.~\ref{fig2}. (a) At the transition point, $\alpha=0.75,$ the numerical results display universal behaviour with all the curves roughly collapsing onto one and consequently matching the threshold value $\epsilon=0.6$ (dashed horizontal line) at the same $\tau.$ (b) The extracted Thouless times $\tau_{\rm Th}$ in the ergodic regime at $\alpha=0.85$ for different system sizes can be matched by the rescaling of the form $\tau \to \tau\exp(bL).$ For this choice of $\epsilon,$ the rescaled values match at $\tau\exp(bL)\approx 2$ (dashed vertical line). All the results shown here (and in Fig.~\ref{fig2}) were calculated using $N_{\rm samples}=1000$ different Hamiltonian realizations.
\label{fig2supp_approach}}
\end{figure*}

{\it Spectral form factor (SFF) implementation.} In the definition and implementation of the SFF $K(\tau)$, we followed the procedure introduced in Ref.~\cite{suntajs_bonca_20a}, which we summarize here for convenience. In Eq.~\eqref{def_Kt}, $\{ \varepsilon_1\le\varepsilon_2\le \cdots\varepsilon_{\cal D} \}$ is the complete set of Hamiltonian eigenvalues after spectral unfolding, $\tau$ is the scaled time, and the average $\langle ... \rangle$ is carried out over different realizations of $\hat H$. We choose the normalization $Z$ such that $K(\tau \gg 1) \simeq 1.$ One can also refer to $K(\tau)$ in Eq.~\eqref{def_Kt} as the unconnected SFF.
We perform the spectral unfolding to eliminate the influence of the local density of states. The operation transforms an ordered set of Hamiltonian eigenvalues $\{E_n\}$ to an ordered set of unfolded eigenvalues $\{\varepsilon_n\},$ for which the local mean level spacing is set to unity at all energy densities. This implies $\mathcal{N}^{-1}\sum_n \delta\varepsilon_n = 1,$ where the averaging is performed in a microcanonical window with $\mathcal{N}$ setting the number of elements in the window, and $\delta\varepsilon_n = \varepsilon_{n+1}-\varepsilon_n.$ After unfolding, the scaled Heisenberg time (or the inverse mean level spacing of the unfolded spectrum) is therefore $\tau_{\rm H}=1$, which is marked by vertical dashed lines in all figures that show $K(\tau)$.
According to the standard unfolding procedure, we introduce the cumulative spectral function $G(E) = \sum_n\Theta\left(E - E_n\right),$ where $\Theta$ is the unit step function. We then smoothen out the stepwise distribution function by fitting a polynomial $\bar{g}_n(E)$ of degree $n$ to $G(E)$ and define the unfolded eigenvalues as $\varepsilon_n = \bar{g}_n(E_n).$ The fitting procedure outlined above can introduce some nonphysical artifacts due to oscillations of the fitting polynomial near the spectral edges, causing the negativity of the unfolded density of states. To prevent this, we only use the states for which the fitting polynomial is non-decreasing and discard the states near the edges where the oscillations occur. In our calculations, we use a fitting polynomial of degree $n=3$ and so the portion of the discarded states is typically negligible.

To avoid the finite size effects due to the spectral edges in Eq.~\eqref{def_Kt}, we use the filtering function $\rho(\varepsilon_n),$ which should not influence the main features of the $K(\tau).$
To achieve this, the filtering function should be ``smooth enough" (e.g.~analytic), symmetric with respect to the center of the unfolded spectrum, and should have a vanishingly small amplitude at the spectral edges. In our calculations, we first perform the unfolding procedure for each disorder realization separately.
Then, we filter each unfolded spectrum using a Gaussian filter, $\rho(\varepsilon_n) = \exp\{\frac{(\varepsilon_n - \bar{\varepsilon})^2}{2(\eta\Gamma)^2}\}$, where $\bar{\varepsilon}$ and $\Gamma^2$ are the average energy and variance at a given Hamiltonian realization, respectively. The dimensionless parameter $\eta$ controls the effective fraction of the eigenstates included in $K(\tau),$ and we set it equal to $\eta=0.5$ in our calculations. This ensures that the overwhelming majority of the eigenstates included in our analysis are the ones that govern the system properties at infinite temperature. We ensure proper normalization, yielding $K(\tau \gg 1) \simeq 1$ in general and $K(\tau) \equiv 1$ for Poisson random spectra, by setting $Z = \langle \sum_n |\rho(\varepsilon_n)|^2\rangle$.

{\it Extraction of the Thouless time.} We compare the numerical results for $K(\tau)$ to the theoretical predictions of the random matrix theory, particularly with the result for the Gaussian Orthogonal ensemble (GOE) for $\tau < 1,$
\begin{equation}\label{eq:def_KGOE}
    K_\mathrm{GOE}(\tau) = 2\tau - \tau\ln(1+2\tau),
\end{equation}
applicable to systems with time-reversal symmetry~\cite{mehta_91}. Using Eq.~\eqref{eq:def_KGOE}, we define the scaled Thouless time $\tau_{\rm Th}$ as the onset time of a universal linear ramp in $K(\tau),$ i.e., for $\tau \geq \tau_{\rm Th},$ we obtain $K(\tau) \simeq K_{\rm GOE}(\tau)$. Due to the noise in $K(\tau)$ upon approaching $K_{\rm GOE}(\tau),$ the extraction of $\tau_{\rm Th}$ is not sharply defined. Below, we briefly discuss the extraction protocol used to obtain the results presented in the main paper.
In extraction of $\tau_{\rm Th},$ we perform the following steps:\\
(i) Each $K(\tau)$ curve is calculated for 5000 times $\tau_i$ in the window $\tau_i \in [1/(2\pi\mathcal{D}), 5]$, with $\tau_i$ equidistant in the logarithmic scale. We then smoothen out the random fluctuations in $K(\tau)$ by performing a running mean such that each new $K(\tau_i)$ is the average over 50 nearest values of $K(\tau_i),$ thereby reducing the final number of data points to 4951.\\
(ii) To analyze the difference between $K(\tau)$ and the GOE prediction given by Eq.~\eqref{eq:def_KGOE}, we define the deviation measure
\begin{equation}\label{eq:def_deviation_measure}
    \Delta K(\tau) = \left|\log_{10} \frac{K(\tau)}{K_{\rm GOE}(\tau)}\right|; \hspace{5mm} \Delta K(\tau_{\rm Th}) = \epsilon.
\end{equation}
We define the scaled Thouless time $\tau_{\rm Th}$ as the time at which $\Delta K(\tau)$ becomes larger than some chosen threshold value $\epsilon$ upon decreasing $\tau$ from the regime $\tau > 1.$ As discussed in Ref.~\cite{suntajs_bonca_20a} in more details, the agreement of $K(\tau)$ with $K_{\rm GOE}(\tau)$ in finite systems is associated with the emergence of noisy data in $\Delta K(\tau),$ with $\Delta K(\tau) \ll 1.$ The chosen $\epsilon$ should then be such that $\Delta K(\tau_{\rm Th})$ is larger than the noise. Throughout our analysis, we used $\epsilon = 0.6,$ which ensured robustness of our results towards random fluctuations at $\tau > 1,$ as shown in Fig.~\ref{fig2supp_approach}.\\
(iii) Upon extraction of the scaled Thouless time $\tau_{\rm Th}$, we calculate the Thouless time in physical units as $t_{\rm Th} = \tau_{\rm Th} t_{\rm H}.$
While the rescaling does not affect the ratio of the Heisenberg and Thouless time which is equal in both scaled and physical units, $t_{\rm H}/t_{\rm Th} = \tau_{\rm H} / \tau_{\rm Th},$ the quantities in Fig.~\ref{fig3} require characteristic times in physical units as an input.

\begin{figure*}[!]
\centering
\includegraphics[width=2\columnwidth]{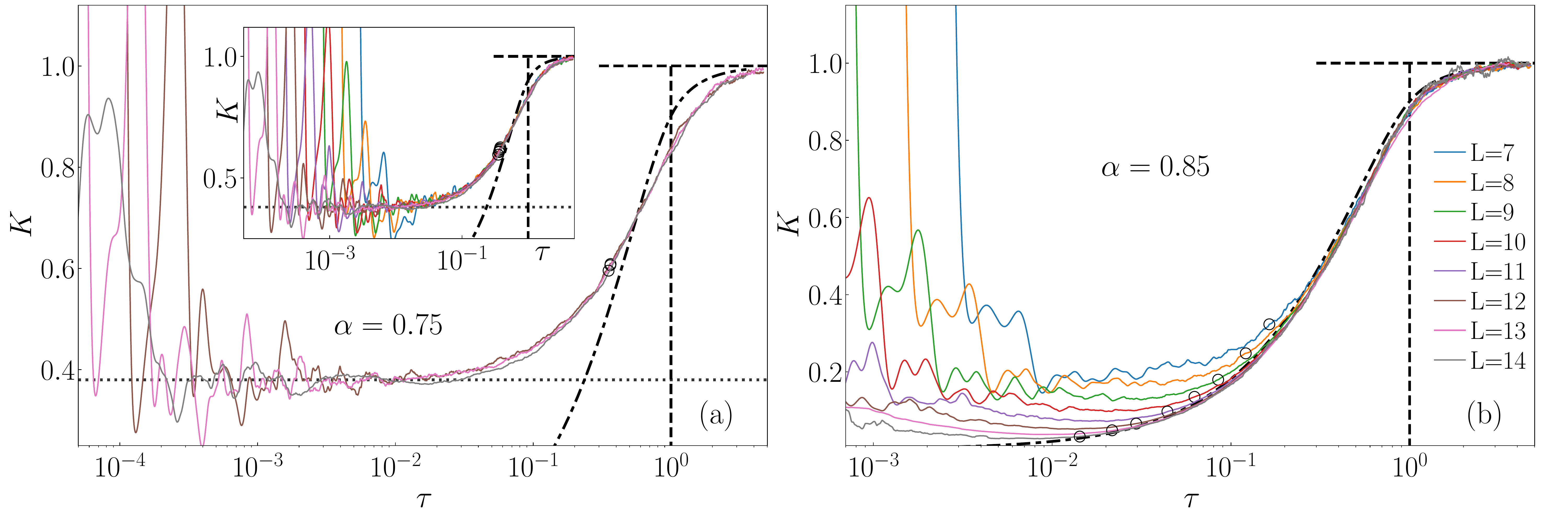}
\caption{A more detailed view into the structure of the $K(\tau)$: (a) at the transition $\alpha=0.75$, and (b) in the ergodic regime at $\alpha=0.85.$ In the former case, a wide nearly $\tau$-independent regime appears at $\tau \ll 1$ prior to the onset of the Thouless time $\tau_{\rm Th}$ (open circles), where $K(\tau) \approx 0.38$ (horizontal dotted line). Similar behavior is observed at the critical point of the 3D Anderson model (see Fig.~\ref{fig2supp_3D_anderson}) where $K(\tau) \approx 0.35$.
Dashed-dotted line denotes $K_{\rm GOE}(\tau)$ and the vertical dashed line denotes $\tau_{\rm H}=1$.
\label{fig2supp}}
\end{figure*}

\begin{figure}[!]
\centering
\includegraphics[width=0.99\columnwidth]{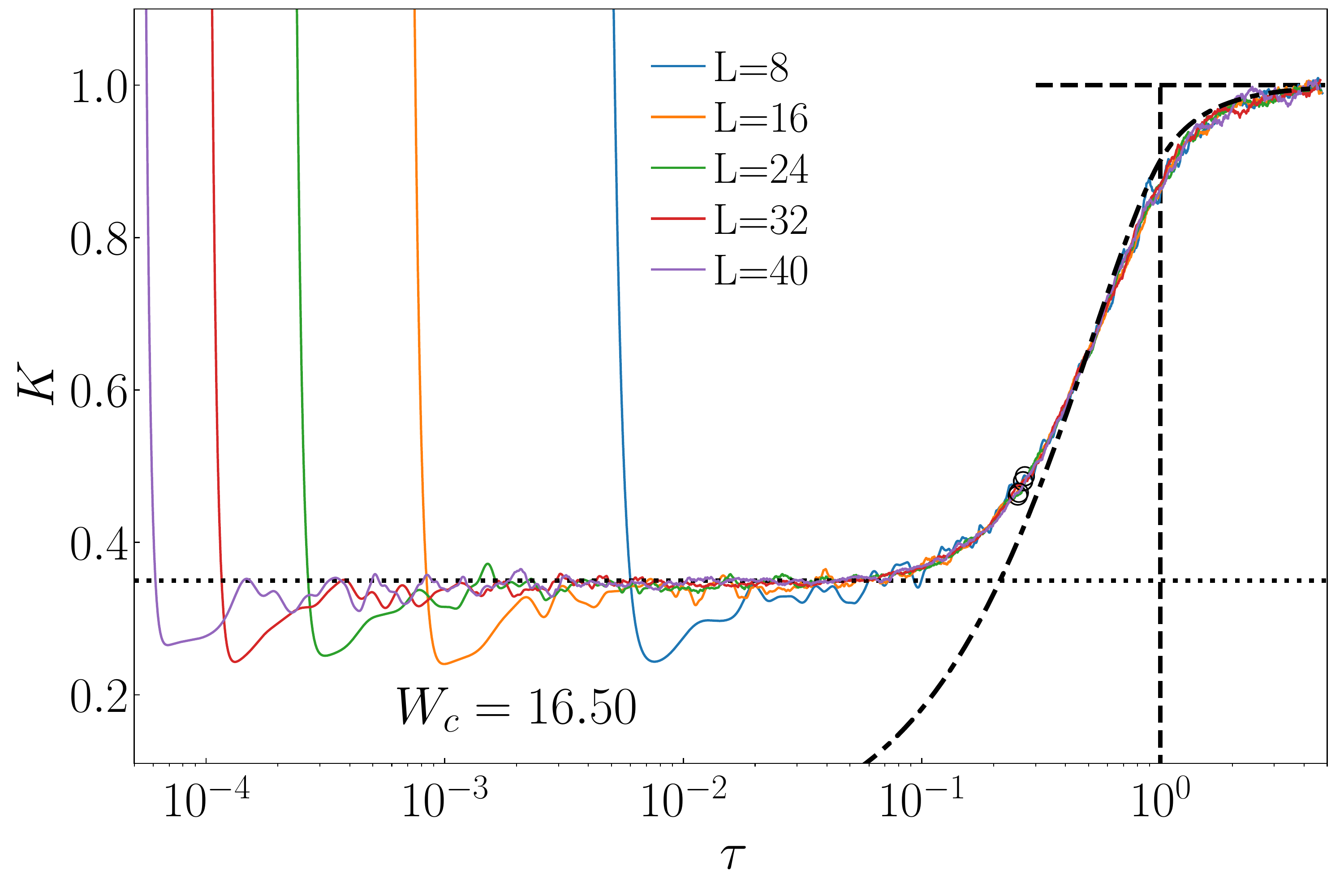}
\caption{Spectral form factor $K(\tau)$ at the critical point of the 3D Anderson model at $W_c=16.50$ for different system sizes $L$, where the total number of sites in a cubic lattice is $L^3$ (see Ref.~\cite{suntajs_prosen_21} for details).
The horizontal dotted line denotes $K(\tau) \approx 0.35$, the dashed-dotted line denotes $K_{\rm GOE}(\tau)$, and the vertical dashed line denotes $\tau_{\rm H}=1$.
\label{fig2supp_3D_anderson}}
\end{figure}

\section{Structure of the SFF at the transition}

In Fig.~\ref{fig2}(a) in the main text, we show the results for $K(\tau)$ at the transition point $\alpha_c,$ highlighting an emerging universal shape similar to the one observed at the localization transition point of the 3D Anderson model~\cite{suntajs_prosen_21}. In Fig.~\ref{fig2supp}(a), we provide a more detailed view of $K(\tau)$ at $\alpha_c,$ displaying plots in a log-linear scale and focusing on a narrower range of values, $K(\tau)\in [0.25, 1.1].$ Prior to the onset of $\tau_{\rm Th}$ at $\tau \ll 1,$ the $K(\tau)$ curves exhibit a broad $\tau$-independent regime where $K(\tau) \approx 0.38,$
which is very close to the value observed at the critical point of the 3D Anderson model, which we show in Fig.~\ref{fig2supp_3D_anderson}. We note that the width of the said regime appears to increase with the system size. Another interesting observation relates to the behavior of $K(\tau)$ at $\tau\approx 1.$ While, based on Fig.~\ref{fig2}, one might conclude that $K(\tau)$ accurately follows $K_{\rm GOE}(\tau)$ for $\tau \geq \tau_{\rm Th},$ a closer look reveals a noticeable difference between the two. This is not the case in the ergodic regime shown in Fig.~\ref{fig2supp}(b).
The origin of the difference may, or may not be, a finite-size effect.
We hence argue that while $\tau_{\rm Th}$ is a time that signals the onset of GOE statistics in the ergodic regime, it is, at least in finite systems close to the transition point and in the nonergodic regime, a characteristic time of the order $O(1)$ obtained by an approximate comparison to $K_{\rm GOE}(\tau)$.

\section{Hybridization condition as a criterion for the EBT} \label{sec:hbd0}

In the main text we used $t_{\rm H}/t_{\rm Th} = {\rm const}$ as a criterion for the EBT.
Here we argue that the hybridization condition yields a criterion for the EBT that is consistent with the former criterion.
The simplest consideration of the hybridization condition refers to the coupling of the farthermost particle $j=L$ to the dot, assuming that the system without that particle is ergodic~\cite{deroeck_huveneers_17}.
It states that the system remains ergodic if the magnitude of the coupling matrix element is larger than the mean level spacing,
\begin{equation}
    g_0 \alpha^{u_L} \langle \hat S_{n_L}^x \hat S_L^x \rangle > \overline{\delta E} \;,
\end{equation}
where $g_0=1$ and $u_L \approx L$.
The matrix element can be estimated using the eigenstate thermalization hypothesis~\cite{deutsch_91, srednicki_94, dalessio_kafri_16} as
$\langle \hat S_{n_L}^x \hat S_L^x \rangle \approx {\cal F}(h) \overline{\delta E}^{1/2}$, where the structure function ${\cal F}$ at the energy $h$ of the spin-flip of the farthermost particle is assumed to be a smooth function of the order $O(1)$ that does not contribute significantly to the scaling analysis.
One can hence introduce the hybridization parameter ${\cal G}_L = \alpha^L/\sqrt{\overline{\delta E}}$, which signals ergodic behavior in the thermodynamic limit if $\lim_{L\to\infty} {\cal G}_L \to \infty$.
Moreover, the square of ${\cal G}_L$ equals
\begin{equation} \label{def_GL2_star}
    {\cal G}_L^2 = \frac{\alpha^{2L}}{\overline{\delta E}} \propto \frac{t_{\rm H}}{t_{\rm Th}^*} \;,
\end{equation}
where $t_{\rm Th}^*$ was introduced in Eq.~(\ref{eq:def_tThstar}) of the main text as an estimate for the Thouless time.
Results in Fig.~\ref{fig3} of the main text show that $t_{\rm Th}^*$ is a reasonable estimate of the Thouless time $t_{\rm Th}$ obtained from the SFF, and hence we conclude that the hybridization condition ${\cal G}_L^2 = {\rm const}$ for the EBT shares the key properties with the criterion $t_{\rm H}/t_{\rm Th} = {\rm const}$ invoked in the main text.

\begin{figure}[!t]
\centering
\includegraphics[width=.99\columnwidth]{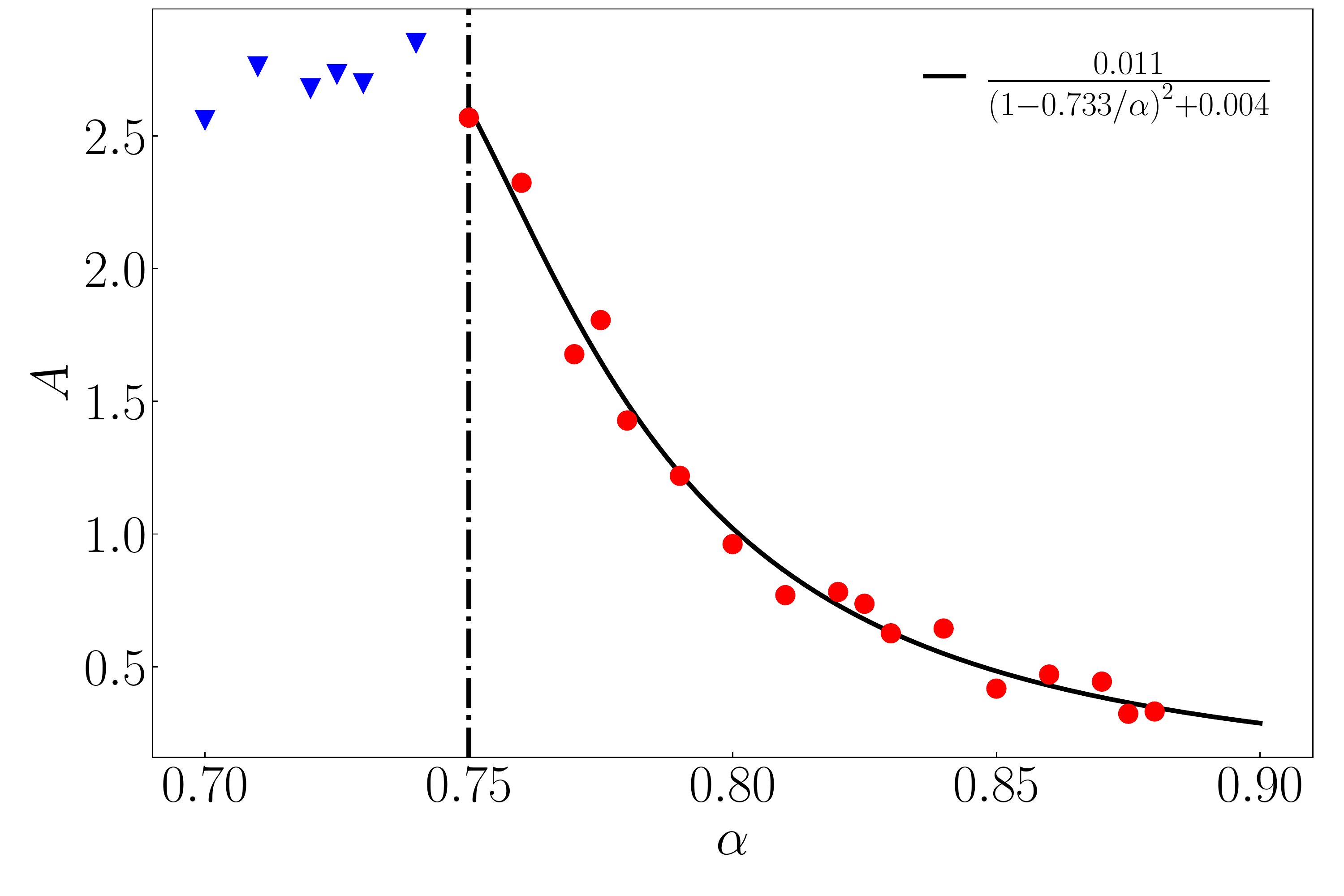}
\caption{Values of the parameter $A$ from Eq.~\eqref{eq:def_GL2} in the main text, as a function of $\alpha$. 
We obtain $A$ by fitting Eq.~\eqref{eq:def_GL2} to the numerical results in Fig.~\ref{fig3}(a) in the nonergodic (blue triangles) and ergodic (red circles) regime.
In the latter, we fit the results using a function $A(\alpha) = A_1/[(1 - \alpha_1/\alpha)^2 + A_2]$ and obtain $\alpha_1=0.733$, $A_1 = 0.011$, and $A_2 = 0.004$.
\label{fig4supp}}
\end{figure}

\section{Corrections to the analytical expression for $t_{\rm Th}$} \label{sec:corrections}

While the analytical arguments presented so far are very powerful and rather simple, they are insufficient to describe the role of subleading contributions to $t_{\rm Th}$, which are inevitably present in finite systems.
For example, in the relation $t_{\rm H}/t_{\rm Th} = A \exp{(L/\xi_{0,1})}$, which is Eq.~(\ref{eq:def_GL2}) in the main text, we had so far no predictions for the $\alpha$ dependence of $A=A(\alpha)$.
In Fig.~\ref{fig4supp} we show numerically extracted values of $A$, obtained by fitting the function in Eq.~(\ref{eq:def_GL2}) to the numerical results in Fig.~\ref{fig3}(a).
On the ergodic side they are well described by a function
$A(\alpha) = A_1/[(1-\alpha_1/\alpha)^2 + A_2]$, where $\alpha_1 = 0.733$ and $A_1 = 0.011$, while the offset $A_2$ is very small, $A_2 = 0.004$.

Here we discuss a possible refinement of analytical arguments that may describe some of the subtle features of numerical results.
We first note that the hybridization parameter ${\cal G}_L$ introduced in the previous section [cf.~Eq.~(\ref{def_GL2_star})] to describe properties of the coupling to the farthermost particle $j=L$ can be generalized to an arbitrary particle $j$ outside the dot.
Starting from the particle that is closest to the dot, $j=1$, we define
\begin{equation}
    {\cal G}_1 = g_1 / \sqrt{\Delta_1} \;,
\end{equation}
where $g_1 = g_0 \alpha$ denotes the coupling and $\Delta_1$ is the many-body level spacing of a system composed of the dot and the particle at $j=1$.
Proceeding iteratively, we define for the next particle, $j=2$,
\begin{equation}
    {\cal G}_2 = g_2 / \sqrt{\Delta_2} \;,\;\;\; g_2 = \alpha g_1,\; \Delta_2 = \Delta_1/2\;,
\end{equation}
and for a general particle at the mean distance $u_j = j$,
\begin{equation}
    {\cal G}_j = (\alpha \sqrt{2})^{j}\, {\cal G}_0 = \left( \frac{\alpha}{\bar\alpha} \right)^j \, {\cal G}_0 \;,
\end{equation}
where $\bar\alpha=1/\sqrt{2}$ and ${\cal G}_0 = g_0/\sqrt{2\Delta_1}$ is a number that does not scale with system size since $g_0=1$ and the dimension of the dot (and hence $\Delta_1$) is fixed.

We note that the sequence of hybridization conditions invoked above is a key ingredient in the description of quantum avalanches~\cite{deroeck_huveneers_17}.
It suggests that if $\alpha>\bar\alpha$, there is a flow towards restoring ergodicity in the thermodynamic limit $j\to\infty$, while in the opposite regime $\alpha<\bar\alpha$ the system remains nonergodic.

We then approximate the global hybridization parameter with the mean over all hybridization parameters,
$\langle {\cal G}\rangle_L = (1/L)\sum_{j=1}^L {\cal G}_j$, which can be calculated exactly.
Its squared value, denoted $\mathcal{G}^2 \equiv \langle \mathcal{G}\rangle_L^2$ should give, as per Sec.~\ref{sec:hbd0}, an estimate of the ratio $t_{\rm H}/t_{\rm Th}$.
It equals, up to a multiplicative factor that we skip for clarity,
\begin{equation} \label{def_G2}
    \mathcal{G}^2 =
    \frac{1}{L^2} \frac{(1- (\frac{\alpha}{\bar\alpha})^L)^2}{(1-\frac{\bar\alpha}{\alpha})^2} =
    \left(\frac{1}{L}\frac{e^{\frac{L}{2\xi_{0}}} - 1}{ e^{-\frac{1}{2\xi_{0}}} - 1}\right)^2\;.
\end{equation}
The last term in Eq.~(\ref{def_G2}) is valid in the ergodic regime $\alpha>\bar\alpha$, while in the nonergodic regime $\alpha<\bar\alpha$ it equals
\begin{equation} \label{def_G2b}
    \mathcal{G}^2 =
    \left(\frac{1}{L}\frac{e^{-\frac{L}{2\xi_{1}}} - 1}{ e^{\frac{1}{2\xi_{1}}} - 1}\right)^2\;.
\end{equation}
The characteristic lengths $\xi_{0}$ and $\xi_1$ are given by Eq.~(\ref{eq:def_xi}).

Equation~(\ref{def_G2}) predicts in the ergodic regime in the limit $L \gg \xi_0 \gg 1$ the scaling ${\cal G}^2 \approx \exp{(L/\xi_0)}(2\xi_0/L)^2$.
While the dominant exponential term $\exp{(L/\xi_0)}$ is hence identical as the one in Eq.~(\ref{eq:def_GL2}), it also contains a polynomial multiplicative factor $(2\xi_0/L)^2$ that is absent in Eq.~(\ref{eq:def_GL2}).
Moreover, the $\alpha-$dependent (and $L-$independent) term in Eq.~(\ref{def_G2}) can be expressed as
$\tilde A(\alpha) = 1/[(1-\bar \alpha/\alpha)^2+A_2]$, where $A_2=0$.
This should be contrasted to the function $A(\alpha)$ that fits the numerical results in Fig.~\ref{fig4supp}, which has $A_2=0.004$.
Both results are indeed very close.

\begin{figure*}[!]
\centering
\includegraphics[width=1.99\columnwidth]{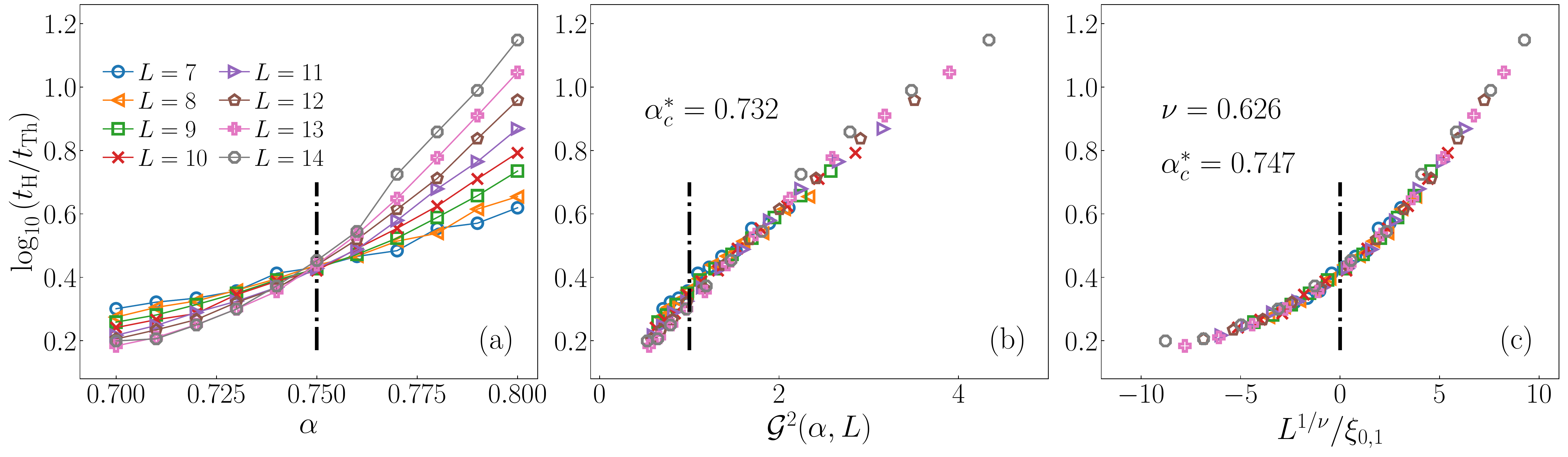}
\caption{(a) Results for $\log_{10}(t_{\rm H}/t_{\rm Th})$ as a function of $\alpha$ in the vicinity of the transition (vertical dot-dashed line denotes $\alpha_c = 0.75$), at different $L$.
(b) Data recast as a function of $\mathcal{G}^2(\alpha,L)$ from Eq.~(\ref{def_G2}).
We sought for an optimal collapse using the cost function minimization procedure~\cite{suntajs_bonca_20}, varying $\alpha_c^*=\bar\alpha$ as a fitting parameter.
We get $\alpha_c^*\approx 0.73$.
(c) Data are recast as a function of $L^{1/\nu}/\xi_{0, 1}$ with $\xi_{0,1}$ defined by Eq.~\eqref{eq:def_xi}. Using the cost function minimization approach with two fitting parameters, we get $\nu\approx 0.63$ and $\alpha^*_c\approx0.75.$
\label{fig5supp}}
\end{figure*}

\begin{figure*}[!]
\centering
\includegraphics[width=1.99\columnwidth]{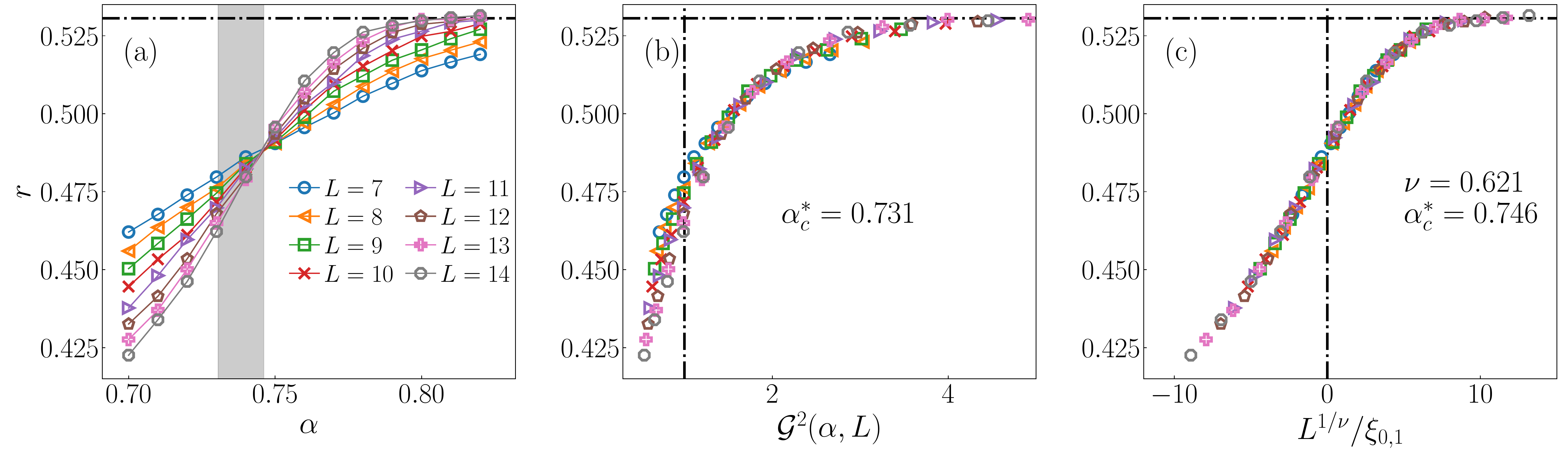}
\caption{(a) Results for the mean level spacing ratio $r$ from Eq.~\eqref{eq:defr}, as a function of $\alpha$ at different $L$.
The scaling collapses in (b) and (c) are obtained as those in Figs.~\ref{fig5supp}(b) and~\ref{fig5supp}(c).
We get $\alpha_c^* \approx 0.73$ in (b) and $\alpha_c^*\approx 0.75$, $\nu\approx0.62,$ in (c).
The interval between both extracted values of $\alpha_c^*$ is colored grey in panel (a).
} \label{fig7supp}
\end{figure*} 

\section{Tests of scaling solutions}

Finally, we test different scenarios for the scaling solutions of the numerical results for the ratio $t_{\rm H}/t_{\rm Th}$ as a function of $\alpha$ and $L$.
The unscaled results are shown in Fig.~\ref{fig4}(b) in the main text and replotted in Fig.~\ref{fig5supp}(a) for convenience.

In Fig.~\ref{fig5supp}(b) we plot $t_{\rm H}/t_{\rm Th}$ versus ${\cal G}^2$ from Eq.~(\ref{def_G2}), for which $\alpha_c^* = \bar \alpha$ is a single fitting parameter of the cost function minimization procedure~\cite{suntajs_bonca_20} for finding the optimal data collapse.
We find $\alpha_c^*\approx 0.73$.
This value is somewhat smaller than $\alpha_c=0.75$ we used throughout the paper, on the other hand, it appears to be consistent with the value $\alpha_1$ obtained in the analysis in Fig.~\ref{fig4supp}.
In Fig.~\ref{fig5supp}(c) we plot $t_{\rm H}/t_{\rm Th}$ versus $L^{1/\nu}/\xi_{0, 1}$, where the two fitting parameters are $\alpha_c^*$ that enters $\xi_{0,1}$, and $\nu$.
We find $\alpha_c^* \approx 0.75$ and $\nu \approx 0.62$.
While the two-parameter scaling collapse in Fig.~\ref{fig5supp}(c) appears to be better than the single-parameter scaling collapse in Fig.~\ref{fig5supp}(b), they both are reasonably good.
However, while analytical arguments that justify the scaling in Fig.~\ref{fig5supp}(b) were presented in Sec.~\ref{sec:corrections}, it is less obvious what should be the arguments to justify the scaling in Fig.~\ref{fig5supp}(c).
More work is needed to clarify these arguments.

In Fig.~\ref{fig7supp}(a), we also show the results for the average level spacing ratio $r$ as a function of $\alpha$ and $L$. To define $r,$ we first define~\cite{oganesyan_huse_07}
\begin{equation}\label{eq:defr}
    \tilde{r}_n = \frac{\min\{\delta E_n, \delta E_{n-1}\}}{\max\{\delta E_n, \delta E_{n-1}\}} = \min\{r_n, r_n^{-1}\}
\end{equation}
for a target eigenstate $\ket{n},$ where $r_n$ is the ratio of consecutive level spacings, $r_n = \delta E_n/\delta E_{n-1},$ and $\delta E_n = E_{n+1} - E_n$ is the level spacing. Since $\{\tilde{r}_n\}$ only assume values from the interval $[0, 1],$ no unfolding is needed to eliminate the influence of finite-size effects through the local density of states. We obtain $r$ by first averaging over $N_{\rm ev} = 500$ eigenstates near the center of the spectrum for each Hamiltonian realization and then over an ensemble of spectra for different Hamiltonian realizations, with $N_{\rm samples} = 500$ for all values of $L.$ The GOE prediction for $r$ is $r_{\rm GOE}\approx 0.5307$~\cite{atas_bogomolny_13}, see the horizontal dashed-dotted lines in Fig.~\ref{fig7supp}. The data used in the calculation of $r$ were obtained using the shift-and-invert method~\cite{pietracaprina2018shift}.
The same quantity was calculated in~\cite{luitz_huveneers_17} for a model very similar to the one in Eq.~(\ref{eq:def_model}).

Rather surprisingly, the $r$-statistics can assume GOE values even in the regime in which the Thouless time $t_{\rm Th}$ scales exponentially with the system size. This is shown in Fig.~\ref{fig7supp}(a), where $r \approx r_{\rm GOE}$ for $\alpha \gtrsim 0.80$ at $L=14.$ As shown in Fig.~\ref{fig3}, the scaling of $t_{\rm Th}$ is exponential in this parameter regime.

In Figs.~\ref{fig7supp}(b) and~\ref{fig7supp}(c) we test the same two scaling collapses as for the ratio $t_{\rm H}/t_{\rm Th}$ in Fig.~\ref{fig5supp}.
In both cases we get a reasonably good scaling collapse, with predictions for the transition point within the interval $\alpha_c^* \in [0.73,0.75]$.
This interval is colored grey in Fig.~\ref{fig7supp}(a).

\end{document}